\documentclass[prd, twocolumn]{revtex4}
\usepackage{amsmath}
\usepackage{graphicx}
\usepackage{dcolumn}
\usepackage{bm}
\usepackage{wrapfig}
\usepackage{epsfig}
\usepackage{breqn}
\usepackage{amssymb}
\usepackage{subfigure}
\usepackage[usenames,dvipsnames]{pstricks}
\usepackage{epsfig}
\usepackage{pst-grad} 
\usepackage{pst-plot} 
 \usepackage[usenames,dvipsnames]{pstricks}
 \usepackage{epsfig}
 \usepackage{hyperref}
 \usepackage{pst-grad} 
 \usepackage{pst-plot} 
 \usepackage{float}

\begin{document}

\title{On Past Singularities in $k=0$ FLRW Cosmologies}
\author{Ikjyot Singh Kohli}
	\email{isk@mathstat.yorku.ca}
	\affiliation{York University - Department of Mathematics and Statistics}
\date{September 6, 2016}                                           


\begin{abstract}
The fundamental singularity theorem of FLRW cosmologies assumes that the matter content in the cosmological model obeys the strong energy condition along with a nonpositive cosmological constant which gives rise to an irrotational geodesic singularity. In this paper, we show that the important case of a spatially flat Friedmann-Lema\^{i}tre-Robertson-Walker universe with barotropic matter obeying only the \emph{weak} energy condition with a nonnegative cosmological constant also contains a past singularity. We accomplish this using topological methods from dynamical systems theory. The methods employed in this paper are sufficiently general that they could be extended to other models to study the existence of past singularities.
\end{abstract}

\maketitle

\section{Introduction}
Some cosmological solutions to the Einstein field equations are understood to begin at a certain cosmic time, which for the sake of simplicity is usually denoted as time $t = 0$, which is usually referred to as the \emph{point of time of the Big Bang} \cite{hervik}. The issue of singularities is important in the context of cosmology because this can possibly give us an answer to whether the Big Bang actually occurred or not. 

If one considers a cosmological constant denoted by $\Lambda$ and denotes the energy density and pressure of the matter in a Friedmann-Lema\^{i}tre-Robertson-Walker (FLRW) universe by $\mu$ and $p$ respectively, then the fundamental singularity theorem for FLRW cosmologies states that if $\Lambda \leq 0, \mu + 3p \geq 0$, and $\mu + p > 0$ (that is, the matter obeys the strong energy condition) in a fluid flow for which there is no acceleration or rotation, then an irrotational geodesic singularity will occur at a finite proper time  \cite{elliscosmo,2007Prama..69...15E, tolmanward, raych1955}. 

With respect to Friedmann-Lema\^{i}tre-Robertson-Walker (FLRW) spacetimes, the concept of initial singularities / big bang singularities has been studied a number of times in the literature in various contexts from classical general relativity to quantum cosmology \cite{1931Natur.128..704L, 1931Natur.127..706L, 1970Natur.228..544H, 1985GReGr..17..397E, 1985GReGr..17..769M, 1986LIACo..26..121S, 1988PNAS...85.7428P, 1991GReGr..23..527H, 1992ApJ...392..385R, 1993RSPSA.443..493N, 1971CMaPh..20...57M, 1994PhLA..188..130C, 1996Prama..47...41A, 2003PhRvD..68f3511S, 2004PhRvD..69l3504C, 2005PhRvD..72f3504S, 2011JETP..112...60K, 2013IJTP...52.2991O, 1994NuPhB.415..497A, 1994CQGra..11.1919A, 1998PhRvD..58f3504M, 2000PhRvD..62d3526C, 2002PhRvD..66b3513I, 2012arXiv1203.1819S, 2012arXiv1207.5303C, 2012Entrp..14.1296G, 2016arXiv160108175D, 1986Ap&SS.123..405B, 1982CaJPh..60..659M, 2007CQGra..24.6243P, 2009PhRvD..79l3509K, 2009PhRvD..80b3501A, 2010PhLB..688..129V, 2012PhRvD..85b3512L, 2012PhRvD..85f4001P, 2014CEJPh..12..123S, 2014PhRvD..90l3538P, 2016IJTP...55...71S, 2007PhLB..656...96B,2004CQGra..21..223E}.  Further, the classic texts \cite{ellis3,waldbook} describe the singularity theorems in significant detail. 

In this paper, we consider a $k=0$ FLRW cosmology with barotropic matter, where $ p = w \mu$, and a positive cosmological constant, $\Lambda > 0$. We demonstrate that assuming only the weak energy condition (WEC), that all such models are past asymptotic to a big bang singularity, thereby extending the classic singularity theorem as described above. Note that, throughout, we use units where $8 \pi G = c = 1$.

\section{The Dynamical Equations}
Following \cite{hervik, ellismac}, we consider an orthonormal frame approach, where one can reduce the Einstein field equations to  a system of first-order, autonomous dynamical equations. Namely, one obtains the Raychaudhuri equation,
\begin{equation}
\label{eq:raych1}
\dot{\theta} + \frac{1}{3}\theta^2 + \frac{1}{2} \left(\mu + 3p \right) - \Lambda = 0,
\end{equation}
the Friedmann equation,
\begin{equation}
\label{eq:fried1}
\frac{1}{3}\theta^2 = \mu + \Lambda,
\end{equation}
and the energy-momentum conservation equation,
\begin{equation}
\label{eq:enmom1}
\dot{\mu} + \theta(\mu + p) = 0.
\end{equation}

Now, the advantage to considering $k=0$ FLRW models, is that one can use the Friedmann equation \eqref{eq:fried1} to eliminate $\mu$ from the Raychaudhuri equation \eqref{eq:raych1}, to obtain a single dynamical equation (after employing a barotropic equation of state, $p = w \mu$):
\begin{equation}
\label{eq:raych2}
\dot{\theta} = -\frac{1}{2} \left(w + 1\right) \left(\theta^2 - 3 \Lambda\right).
\end{equation}

The big bang singularity occurs for when $\theta = \infty$. It is of interest to determine whether $\theta \to \infty$ is indeed a past asymptotic state of such a model. We wish to compactify this infinity by employing the following coordinate transformation, which takes place in the phase space of our ordinary differential equation. Namely, we let
\begin{equation}
\label{eq:transform1}
X = \frac{1}{1 + \exp(-\theta)}.
\end{equation}
Therefore, we have that
\begin{equation}
\lim_{\theta \to \infty} X  = 1.
\end{equation}
Therefore, the singularity $\theta = \infty$ appears as $X = 1$ in our transformed variable.

Further, under the transformation Eq. \eqref{eq:transform1}, the Raychauhduri equation \eqref{eq:raych2} takes the form
\begin{equation}
\label{eq:raych3}
\dot{X} = -\frac{1}{2}(1+w)(X-1)X \left[3\Lambda - \log^2 \left(-1 + \frac{1}{X}\right)\right].
\end{equation}

\section{The Singularity as a Source of Raychaudhuri's Equation}
With the Raychaudhuri equation \eqref{eq:raych3} in hand, we are in a position to describe its fixed points and behaviour with respect to these points. In actuality, we are only interested in the point $X=1$ which corresponds to $\theta = \infty$ in the original coordinates. To prove the existence of a big bang singularity, our goal is to show that the point $X=1$ is a past asymptotic state of Eq. \eqref{eq:raych3}.

The first order of business is to determine whether $X=1$ is indeed a fixed point of Eq. \eqref{eq:raych3}. Indeed, there may be some objection to this because the $\log$ function on the right hand side of Eq. \eqref{eq:raych3} blows up for $X = 1$. However, note that
\begin{equation}
\lim_{X \to 1}  -\frac{1}{2}(1+w)(X-1)X \left[3\Lambda - \log^2 \left(-1 + \frac{1}{X}\right)\right] = 0.
\end{equation}
Therefore, $X=1$ is indeed a fixed point of Eq. \eqref{eq:raych3}. We have compactified the big bang singularity to being a fixed point of our transformed equation. 

What remains to be done is to prove that $X=1$ is a past asymptotic state of the dynamics represented by Eq. \eqref{eq:raych3}. The standard methodology is to first determine local behaviour, that is, whether $X=1$ is a source of the system by analyzing the the sign of the first derivative of the right-hand-side of Eq. \eqref{eq:raych3} at $X=1$. However, this cannot be done in this case. If we denote the right-hand-side of Eq. \eqref{eq:raych3} by $f(X)$, then $f'(X)$ is given by
\begin{dmath}
f'(X) = -\frac{1}{2} (w+1) \left[3 \Lambda  (2 X-1)+(1-2 X) \log ^2\left(\frac{1}{X}-1\right)-2 \log \left(\frac{1}{X}-1\right)\right],
\end{dmath}
which is not defined at $X=1$. Further, 
\begin{equation}
\lim_{X \to 1} f'(X) = -\infty.
\end{equation}
Therefore, a local stability analysis as suggested in \cite{ellis}, for example, will not work in this case.

We therefore make use of Chetaev's instability theorem to show that $X=1$ is unstable, that is, it is a past state of Eq. \eqref{eq:raych3}. Following \cite{arnolddyn}, we note that a differentiable function $f$ is called a \emph{Chetaev function} for a singular point $x_{0}$ of a vector field $v$ if it satisfies the following conditions: the function $f$ is defined on a domain $W$ whose boundary contains $x_{0}$, then $f(x) \to 0$ as $x \to x_{0}, \quad x \in W$; $f > 0$ and $\dot{f} > 0$ everywhere in $W$. Then \emph{Chetaev's instability theorem} states that a singular point for which a Chetaev function exists is unstable.

Motivated by this theorem, let us define a function $Z$, such that
\begin{equation}
Z = -\log X,
\end{equation}
and a domain $W$ such that
\begin{equation}
W = \left\{\frac{1}{1 + \exp \left(-\sqrt{3} \sqrt{\Lambda}\right)} < X  < 1\right\}.
\end{equation}
Clearly, the boundary of $W$ is given by
\begin{equation}
\partial W = \left\{X = \frac{1}{1 + \exp \left(-\sqrt{3} \sqrt{\Lambda}\right)}\right\} \cup \left\{X = 1 \right\},
\end{equation}
which importantly contains the fixed point $X = 1$. 

To apply Chetaev's theorem, first note that
\begin{equation}
\lim_{X \to 1} Z = \lim_{X \to 1} -\log X = 0.
\end{equation}
Further, the function $Z$, and its derivative, $\dot{Z} = -\dot{X}/X$ are clearly strictly positive in $W$ (which can be seen using Eq. \eqref{eq:raych3}) on the condition that $-1 < w \leq 1$ and $\Lambda > 0$. For a barotropic equation of state the condition $w > -1$ is just the weak energy condition. The condition $\Lambda > 0$ indicates that we must have a positive cosmological constant. 

Therefore, we have just shown that for a positive cosmological constant and assuming that the matter distribution in our model satisfies the weak energy condition, $X=1$ is indeed a source of Eq. \eqref{eq:raych3}. That is, there exists a one-dimensional unstable manifold of $X=1$ that is tangent to the one-dimensional unstable subspace at $X=1$ such that all orbits in the unstable manifold are asymptotic to $X=1$ as $t \to -\infty$. 

What remains to be shown is whether $X=1$ is indeed asymptotically unstable. To accomplish this, we will use the LaSalle invariance principle \cite{ellis}, but modified for $\alpha$-limit sets. This reads as follows. Consider a dynamical system $x' = f(x)$ on $\mathbb{R}^{n}$, with flow  $\phi_{t}$. Let $S$ be a closed, bounded, and negatively invariant set of $\phi_{t}$, and let $Z$ be a $C^{1}$ monotone function, where $\dot{Z} \leq 0$. Then, for all $x_{0}$ in $S$, $\alpha(x_{0}) \subseteq \{x \in S | \dot{Z} = 0\}$.  Let $Z_{1} = X$. Then, $Z_1$ is clearly monotone for $X=1$ which is a local source by the preceding analysis, and is therefore a negatively invariant closed set. Therefore, applying the LaSalle invariance principle, we find that $\alpha(x) = \{X=1\}$. We therefore conclude that $X=1$ is asymptotically unstable.

In other words, since $X=1$ represents $\theta = \infty$ in the original variable choice, we have just shown that \emph{a $k=0$ FLRW model with a positive cosmological constant and barotropic matter obeying only the weak energy condition is past asymptotic to a big bang singularity.}

\section{Conclusions}
In this brief paper, we  considered the dynamics of a spatially flat FLRW spacetime with a positive cosmological constant and matter obeying a barotropic equation of state. By performing a coordinate transformation in the configuration space on the Raychaudhuri equation in order to compactify the big bang singularity to a finite point. The fundamental singularity theorem of FLRW cosmologies assumes that the matter content in the cosmological model obeys the strong energy condition along with a nonpositive cosmological constant which gives rise to an irrotational geodesic singularity. We showed that a spatially flat Friedmann-Lema\^{i}tre-Robertson-Walker universe with barotropic matter obeying only the \emph{weak} energy condition with a \emph{nonnegative} cosmological constant also contains a past singularity. This was demonstrated by first showing that the big bang singularity point was locally unstable using Chetaev's instability theorem. We then used the LaSalle invariance principle to show that this point is in fact an asymptotically unstable equilibrium point of Raychaudhuri's equation as applied to the model under consideration.

\section{Acknowledgements}
The author would like to thank Michael C. Haslam, Frits Veerman and Juli\'{a}n Aguirre for helpful discussions regarding fixed points of dynamical systems.

\bibliographystyle{ieeetr} 
\bibliography{sources}

\end{document}